# Multipole expansions in quantum radiation theory

M. Ya. Agre

*National University of "Kyiv-Mohyla Academy", 04655 Kyiv, Ukraine*
E-mail: magrik@ukr.net

With the help of mathematical technique of irreducible tensors the multipole expansion for the probability amplitude of spontaneous radiation of a quantum system is derived. It is shown that the found series represents the total radiation amplitude in the form of the sum of radiation amplitudes of electric and magnetic $2^l$-pole ($l$ = 1, 2, 3,…) photons. All information about the radiating system is contained in the coefficients of the series which are the irreducible tensors being determined by the current density of transition. The expansion can be used for solving different problems that arise in studying electromagnetic field-quantum system interaction both in long-wave approximation and outside its framework.

PACS number(s): 31.15.ag, 32.70.-n, 32.90.+a

## I. INTRODUCTION

Multipole expansions in classical electrodynamics are the seriess for potentials or field strengths, where the coefficients are the irreducible tensors dependent on the charge or current distributions of the system that is the source of the field. In every term of the multipole series the irreducible tensors have been convoled in scalars (for the scalar potential) or vectors (for the vector potential or field strength) with the irreducible tensors specifying the corresponding fields (called the multipole fields [1]).

The expression for multipole expansion in electrostatics can be found in any textbook on classical electrodynamics (see, e.g., [2]). In that case the irreducible tensors specifying the source are the well known electric multipole moments of the charge system. The structure of multipole expansion in magnetostatics is more complicated owing to the vector nature of the magnetic field potential. And no wonder that one can only find the first term of the expansion in the textbook which is determined by the field of magnetic dipole and contains the magnetic (magnetic dipole) moment of the the current system as the coefficient. The total multipole expansion for magnetic field of a current system containing magnetic multipole moments has been recently derived by us [3].

The different forms of multipole expansions in classical radiation theory are known [1,4]. Note that the expansion for the potential of radiation field given by Dubovik and Cheshkov [4] is considerably more convenient for applications. It has allowed to accomplish the correct analysis of long-wave approximation in the radiation theory and to take into account the contribution of toroid multipole moments arising in higher orders of the approximation [4].

Multipole expansion has also to work effectively in the quantum radiation theory, especially under consideration of transitions between the states with definite values of angular momentum of the radiating system when the angular momentum selection rules take place. We needed such expansion in theoretical study of multipole radiation of light by oriented and aligned (i.e., spin-polarized) atoms in long-wave approximation[1]. The known multipole expansion for the probability amplitude of a quantum transition accompanied by radiation of photon with definite polarization and direction of motion is based on representation of function $\mathbf{e}_\pm \exp(i\mathbf{k}\cdot\mathbf{r})$, where $\mathbf{e}_\pm$ is the vector of right(left)-hand circular polarization and $\mathbf{k}$ is the wave vector (i.e., the momentum of the photon in the units of $\hbar$), in the form of the series in the spherical vectors or the fields of electric and magnetic multipoles (see, e.g., [1]). The coefficients of the series are proportional to Wigner's $D^L_{M,\pm 1}(\varphi,\theta,0)$-function, where $\varphi$ and $\theta$ are the azimuthal and polar angles specifying the direction of the vector $\mathbf{k}$. In the interesting for us case of multipole radiation by spin-polarized atoms this expansion led to cumbersome formulas that essentially hampered the study. That is why we were forced to seek for other form of multipole expansion. As a result, the multipole series had been derived which is highly efficient, in our opinion, in solution of different problems of quantum radiation theory both in long-wave approximation and outside its framework.

Note that analyzing the long-wave approximation in quantum radiation theory the authors of known textbooks on quantum electrodynamics [6,7] do at all without the multipole expansion for the amplitude of photon radiation with definite momentum and polarization. On the first stage they limit themselves to derive the total radiation probability of electric $2^l$-pole ($El$) or magnetic $2^l$-pole ($Ml$) photons. Considering further the multipole radiation of given direction and polarization the authors [6] actually construct in artificial way the corresponding term of the multipole expansion, where the Wigner's D-function arises again. In such way one cannot furthermore take into account the possible interference of $Ml$ and $El'$, $l'=l+1$,

---
[1] The results of the study have been presented at ECAMP11 [5].

amplitudes which have, generally speaking, the same order in the long-wave approximation.

The paper is organized as follows. In Section II the mathematical technique for deriving multipole expansion in quantum radiation theory is stated and the multipole series for the radiation amplitude is found. Several equivalent forms of the multipole series are given in Sec. III. The possible applications of the derived expansion are discussed in the final section of the paper.

## II. MULTIPOLE EXPANSION FOR THE AMPLITUDE OF SPONTANEOUS RADIATION

The probability of spontaneous radiation per unit time of a photon with frequency $\omega$, unit polarization vector $\mathbf{e}$ and wave vector $\mathbf{k}$ in spatial angle element $d\Omega$ is determined by the well known expression (see, e.g., [6])

$$dW_{fi} = \frac{\omega}{2\pi\hbar c^3}|V_{fi}|^2 d\Omega,$$

where

$$V_{fi} = \mathbf{e}^* \cdot \int \mathbf{j}_{fi}(\mathbf{r})\exp(-i\mathbf{k}\cdot\mathbf{r})d\mathbf{r}. \quad (1)$$

The radiating system passes with that from the initial state $|i\rangle$ into the final state $|f\rangle$. The value $V_{fi}$ (1), where $\mathbf{j}_{fi}$ is the current density vector of the transition, we shall call the probability amplitude of spontaneous radiation and shall seek the multipole expansion just for it. Although the explicit structure of the current density vector is not used in the derivation of multipole expansion, the corresponding formulas are given in Appendix A to complete the statement.

Let us introduce the vector

$$\mathbf{I}_{fi} = \int \mathbf{j}_{fi}(\mathbf{r})\exp(-i\mathbf{k}\cdot\mathbf{r})d\mathbf{r} \quad (2)$$

and write the radiation amplitude (1) in the form of scalar product:

$$V_{fi} = \mathbf{e}^* \cdot \mathbf{I}_{fi}. \quad (3)$$

Let us also use the known expansion [8] for the exponential function

$$\exp(i\mathbf{k}\cdot\mathbf{r}) = 4\pi\sum_{l=0}^{\infty}\sum_{m=-l}^{l}i^l g_l(kr)Y_{lm}(\hat{\mathbf{r}})Y_{lm}^*(\hat{\mathbf{k}}),$$

where $\hat{\mathbf{r}} = \mathbf{r}/r$ and $\hat{\mathbf{k}} = \mathbf{k}/k$ are the unit vectors,

$$g_l(x) = \sqrt{\frac{\pi}{2x}}J_{l+1/2}(x), \quad (4)$$

$J_\nu(x)$ is Bessel function and $Y_{lm}$ is spherical function. Substituting this expansion in equation (2) gives the following series for $\mathbf{I}_{fi}$:

$$\mathbf{I}_{fi} = 4\pi\sum_{l,m}(-i)^l \int g_l(kr)\mathbf{j}_{fi}(\mathbf{r})Y_{lm}^*(\hat{\mathbf{r}})d\mathbf{r}Y_{lm}(\hat{\mathbf{k}}). \quad (5)$$

To derive the multipole expansion we take advantage of mathematical technique of angular momentum algebra (i.e., the technique of irreducible tensors which is based on the theory of irreducible representation of the rotation group) and separate out in the series (5) and respectively in the expansion of the radiation amplitude (1), (3) the irreducible tensors being determined by the current density of transition. We use further a number of standard definitions and designations of the technique [9] which we give here for the convenience. The spherical components $a_m$, $m = 0, \pm 1$, of arbitrary vector $\mathbf{a}$ forming the first rank irreducible tensor are expressed through its Cartesian components as follows:

$$a_0 = a_z, \quad a_{\pm 1} = \mp\frac{1}{\sqrt{2}}(a_x \pm ia_y).$$

One can compose of two irreducible tensors $A_{lm}$ and $B_{l'm'}$ having respectively the ranks $l$ and $l'$ the irreducible tensors (the tensor products of the irreducible tensors) of the ranks

$$L = |l-l'|, |l-l'|+1, \ldots, l+l'$$

after the rule

$$\{A_l \otimes B_{l'}\}_{LM} = \sum_{m,m'} C_{lml'm'}^{LM} A_{lm} B_{l'm'}, \quad (6)$$

where $C_{lml'm'}^{LM}$ is the Clebsh-Gordan coefficient. If $l = l'$, the scalar (the zero rank irreducible tensor) is also composed. The scalar is proportional to the scalar product of two tensors that will be designated by parentheses:

$$(A_l \cdot B_l) = \sum_m (-1)^m A_{l,-m} B_{lm}$$
$$= (-1)^l\sqrt{2l+1}\{A_l \otimes B_l\}_{00}. \quad (7)$$

In particular, the zero rank tensor composed of two vectors is proportional to the ordinary scalar product of the vectors:

$$\{\mathbf{a} \otimes \mathbf{b}\}_{00} = -\frac{1}{\sqrt{3}} \mathbf{a} \cdot \mathbf{b}. \qquad (8)$$

With the help of the technique analogous of that, used by us in derivation of multipole expansion in magnetostatics [3], one can separate out the irreducible tensors dependent on the current density of transition in the expression (5). To do it we note that the spherical components of the vector $\mathbf{I}_{fi}$ (5) contain the construction of the following form:

$$\left(j_{fi}\right)_m \sum_{m'} (-1)^{m'} Y_{l,-m'}(\hat{\mathbf{r}}) Y_{lm'}(\hat{\mathbf{k}}). \qquad (9)$$

Here $\left(j_{fi}\right)_m$ is the spherical component of the current density vector of transition and the identity $Y_{lm}^* = (-1)^m Y_{l,-m}$ for the spherical function has been taken into consideration. Taking also into account the definition of the scalar product of two irreducible tensors (7), one can write the expression (9) in the form of the tensor product in the designations (6),

$$(-1)^l \sqrt{2l+1} \left\{ \mathbf{j}_{fi} \otimes \left\{ Y_l(\hat{\mathbf{r}}) \otimes Y_l(\hat{\mathbf{k}}) \right\}_0 \right\}_{1m},$$

and it is sufficient to change by standard way [9] the coupling scheme of angular momenta in this expression to achieve our object. In given case the tensor product is simple, thus one can do without 6*j*-symbols changing the coupling scheme. Really, with the help of relation, which is inverse to (6), we find that

$$\left(j_{fi}\right)_m Y_{lm'}^*(\hat{\mathbf{r}}) = (-1)^{m'} \left(j_{fi}\right)_m Y_{l,-m'}(\hat{\mathbf{r}})$$
$$= (-1)^{m'} \sum_{L,M} C_{1ml,-m'}^{LM} \left\{ \mathbf{j}_{fi} \otimes Y_l(\hat{\mathbf{r}}) \right\}_{LM}.$$

Making further use of symmetry properties of Clebsh-Gordan coefficients leading to the identity

$$C_{1ml,-m'}^{LM} = (-1)^{L-1-m'} \sqrt{\frac{2L+1}{3}} C_{LMlm'}^{1m},$$

one reduces the expression (9) to the form

$$\left(j_{fi}\right)_m \sum_{m'} Y_{lm'}^*(\hat{\mathbf{r}}) Y_{lm'}(\hat{\mathbf{k}})$$
$$= \sum_L (-1)^{L-1} \sqrt{\frac{2L+1}{3}} \left\{ \left\{ \mathbf{j}_{fi} \otimes Y_l(\hat{\mathbf{r}}) \right\}_L \otimes Y_l(\hat{\mathbf{k}}) \right\}_{1m}. \qquad (10)$$

Substituting the identity (10) in (5) and taking into consideration the relation

$$\left\{ \mathbf{j}_{fi} \otimes Y_l(\hat{\mathbf{r}}) \right\}_{LM} = (-1)^{l+1-L} \left\{ Y_l(\hat{\mathbf{r}}) \otimes \mathbf{j}_{fi} \right\}_{LM},$$

we find as a result the following series for the probability amplitude of spontaneous radiation $V_{fi}$ (3):

$$V_{fi} = \sum_{l=0}^{\infty} \sum_{L} (-1)^l \sqrt{\frac{2L+1}{3}} \left( \left\{ A_L^{(l)} \otimes Y_l(\hat{\mathbf{k}}) \right\}_1 \cdot \mathbf{e}^* \right). \qquad (11)$$

Here the irreducible tensors

$$A_{LM}^{(l)} = 4\pi (-i)^l \int \left\{ g_l(kr) Y_l(\hat{\mathbf{r}}) \otimes \mathbf{j}_{fi}(\mathbf{r}) \right\}_{LM} d\mathbf{r} \qquad (12)$$

are introduced which are determined by the current density of transition and thus specify the radiating quantum system. Note also that the index $L$ in the internal sum (11) satisfies the triangle condition $|l-1| \le L \le l+1$ (i.e., $L = l, l \pm 1$ if $l \ne 0$ and $L = 1$ if $l = 0$) in accordance with angular momenta addition rule.

The irreducible tensors $A_{LM}^{(l)}$ (12) can also be expressed in terms of spherical vectors

$$\mathbf{Y}_{LlM}(\hat{\mathbf{r}}) = \sum_{m_1 m_2} C_{lm_1 1 m_2 0}^{LM} Y_{lm_1}(\hat{\mathbf{r}}) \mathbf{e}_{m_2} \qquad (13)$$

[1,7]. Here the unit vectors $\mathbf{e}_m$, $m = 0, \pm 1$, are determined in terms of unit vectors $\mathbf{e}_x, \mathbf{e}_y$ and $\mathbf{e}_z$ of Cartesian coordinate system:

$$\mathbf{e}_0 = \mathbf{e}_z, \quad \mathbf{e}_{\pm 1} = \mp \frac{1}{\sqrt{2}} \left( \mathbf{e}_x \pm i \mathbf{e}_y \right).$$

Taking into account the expression $a_m = \mathbf{a} \cdot \mathbf{e}_m$ for the spherical components of the vector $\mathbf{a}$, we find the irreducible tensor $A_{LM}^{(l)}$ (12) in equivalent form

$$A_{LM}^{(l)} = 4\pi (-i)^l \int \left( g_l(kr) \mathbf{Y}_{LlM}(\hat{\mathbf{r}}) \cdot \mathbf{j}_{fi}(\mathbf{r}) \right) d\mathbf{r}. \qquad (14)$$

The last step to derive the multipole expansion consists in changing the coupling scheme of angular momenta in the scalar products which enter the series (11). Making use for that the formula (B2) of Appendix B, we derive from (11) the multipole expansion of the radiation amplitude (1):

$$V_{fi} = \sum_{l=0}^{\infty} \sum_{L=|l-1|}^{l+1} \left( A_L^{(l)} \cdot \left\{ \mathbf{e}^* \otimes Y_l(\hat{\mathbf{k}}) \right\}_L \right). \qquad (15)$$

Let us also give the multipole series for the value, complex conjugate of the radiation amplitude (1):

$$V_{fi}^* = \mathbf{e} \cdot \int \mathbf{j}_{if}(\mathbf{r}) \exp(i \mathbf{k} \cdot \mathbf{r}) d\mathbf{r},$$

where the relation $\mathbf{j}_{fi}^* = \mathbf{j}_{if}$ (see Appendix A) has been taken into consideration. Comparing this expression with (1) shows that the multipole expansion for $V_{fi}^*$ can be found from (15) by substitutions of $\mathbf{e}$, $-\hat{\mathbf{k}}$ and $\mathbf{j}_{if}$ for $\mathbf{e}^*$, $\hat{\mathbf{k}}$ and $\mathbf{j}_{fi}$ respectively. Considering also that $Y_{lm}(-\hat{\mathbf{k}}) = (-1)^l Y_{lm}(\hat{\mathbf{k}})$, one derives the multipole expansion in the following form:

$$V_{fi}^* = \sum_{l=0}^{\infty} \sum_{L=|l-1|}^{l+1} (-1)^l \left( B_L^{(l)} \cdot \{\mathbf{e} \otimes Y_l(\hat{\mathbf{k}})\}_L \right),$$

where the irreducible tensors $B_{LM}^{(l)}$ are determined by the equation (12) or (14) with the substitution of $\mathbf{j}_{if}$ for $\mathbf{j}_{fi}$ (i.e., with the rearrangement of the indices of the current density vector).

## III. MULTIPOLE EXPANSION AND RADIATION OF $El$ AND $Ml$-PHOTONS

It is known that the angular dependence of the photon wave function in the states of electric ($El$-type (the angular momentum quantum number is equal to $l$ and parity is equal to $(-1)^l$) is determined in momentum representation by linear combination of two spherical vectors (13):

$$\mathbf{Y}_{lm}^{(E)}(\hat{\mathbf{k}}) = \sqrt{\frac{l+1}{2l+1}} \mathbf{Y}_{l,l-1,m}(\hat{\mathbf{k}}) + \sqrt{\frac{l}{2l+1}} \mathbf{Y}_{l,l+1,m}(\hat{\mathbf{k}}), \quad (16)$$

whereas for the states of magnetic ($Ml$)-type (the angular momentum quantum number is equal to $l$ and parity is equal to $(-1)^{l+1}$) the angular part of the photon wave function is determined by spherical vector alone:

$$\mathbf{Y}_{lm}^{(M)}(\hat{\mathbf{k}}) = \mathbf{Y}_{llm}(\hat{\mathbf{k}}) \quad (17)$$

[6,7]. Let us prove that the series (15) found by us can be considered as an expansion of the amplitude of a photon radiation in direction $\hat{\mathbf{k}}$ with polarization vector $\mathbf{e}$ in the corresponding radiation amplitudes of electric-type and magnetic-type photons.

Note first of all that the term with $l=1$ and $L=0$ is absent in the series (15). Really, the spherical function $Y_{1m}(\hat{\mathbf{k}})$ is proportional to the spherical components of the unit vector $\hat{\mathbf{k}} = \mathbf{k}/k$,

$$Y_{1m}(\hat{\mathbf{k}}) = \sqrt{\frac{3}{4\pi}} \hat{k}_m.$$

Therefore, taking the identity (8) into consideration, we find that

$$\{\mathbf{e}^* \otimes Y_1(\hat{\mathbf{k}})\}_{00} = -\frac{1}{\sqrt{4\pi}} \mathbf{e}^* \cdot \hat{\mathbf{k}} = 0$$

in accordance with the transversality condition for the radiation field. The pointed reason allows us to write the expansion (15) in the following form:

$$V_{fi} = \sum_{L=1}^{\infty} \sum_{l=L-1}^{L+1} \left( A_L^{(l)} \cdot \{\mathbf{e}^* \otimes Y_l(\hat{\mathbf{k}})\}_L \right). \quad (18)$$

The irreducible tensors composed of the polarization vector and the spherical function in (18) (see definition (6)) can be expressed in terms of the spherical vectors (13):

$$\{\mathbf{e}^* \otimes Y_l(\hat{\mathbf{k}})\}_{Lm} = (-1)^{l+1-L} \{Y_l(\hat{\mathbf{k}}) \otimes \mathbf{e}^*\}_{Lm}$$
$$= (-1)^{l+1-L} \mathbf{e}^* \cdot \mathbf{Y}_{Llm}(\hat{\mathbf{k}}). \quad (19)$$

The scalars of three types enter the expansion (18) of the radiation amplitude. Let us introduce the following designation for one of them:

$$V_l^{(M)} \equiv \left( A_l^{(l)} \cdot \{\mathbf{e}^* \otimes Y_l(\hat{\mathbf{k}})\}_l \right)$$
$$= -\left( A_l^{(l)} \cdot \left( \mathbf{e}^* \cdot \mathbf{Y}_{ll}(\hat{\mathbf{k}}) \right) \right), \quad (20)$$

where

$$\left( A_l^{(l)} \cdot \left( \mathbf{e}^* \cdot \mathbf{Y}_{ll}(\hat{\mathbf{k}}) \right) \right) = \sum_m (-1)^m A_{l,-m}^{(l)} \left( \mathbf{e}^* \cdot \mathbf{Y}_{llm}(\hat{\mathbf{k}}) \right),$$

and the identity (19) and the definition of scalar product (7) are taken into account. The spherical vector entering the expression $V_l^{(M)}$ (20) is directly connected with the photon wave function in the state of $Ml$-type, thus one can write using the designation (17) that

$$V_l^{(M)} = -\left( A_l^{(l)} \cdot \left( \mathbf{e}^* \cdot \mathbf{Y}_l^{(M)}(\hat{\mathbf{k}}) \right) \right). \quad (21)$$

The scalar product of the polarization vector and the spherical vector (17) in equation (21) is proportional to projection of the photon $Ml$-state onto the state with definite polarization and momentum that allows of interpreting $V_l^{(M)}$ as the contribution of magnetic $2^l$-pole photon to the total radiation amplitude [6].

Let us further write the expansion of the radiation amplitude (18) in the form

$$V_{fi} = \sum_{l=1}^{\infty} \left( V_l^{(M)} + V_l^{(E)} \right) \qquad (22)$$

and ascertain the physical meaning of the amplitude $V_l^{(E)}$,

$$\begin{aligned} V_l^{(E)} &= \left( A_l^{(l-1)} \cdot \left\{ \mathbf{e}^* \otimes Y_{l-1}(\hat{\mathbf{k}}) \right\}_l \right) \\ &+ \left( A_l^{(l+1)} \cdot \left\{ \mathbf{e}^* \otimes Y_{l+1}(\hat{\mathbf{k}}) \right\}_l \right) \\ &= \left( A_l^{(l-1)} \cdot \left( \mathbf{e}^* \cdot \mathbf{Y}_{l,l+1}(\hat{\mathbf{k}}) \right) \right) \\ &+ \left( A_l^{(l+1)} \cdot \left( \mathbf{e}^* \cdot \mathbf{Y}_{l,l+1}(\hat{\mathbf{k}}) \right) \right). \end{aligned} \qquad (23)$$

For that purpose, making use of (16) and the expansion of the longitudinal vector $\hat{\mathbf{k}} Y_{lm}(\hat{\mathbf{k}})$ in the spherical vectors,

$$\hat{\mathbf{k}} Y_{lm}(\hat{\mathbf{k}}) = \sqrt{\frac{l}{2l+1}} \mathbf{Y}_{l,l-1,m}(\hat{\mathbf{k}}) - \sqrt{\frac{l+1}{2l+1}} \mathbf{Y}_{l,l+1,m}(\hat{\mathbf{k}})$$

[7], we express the spherical vectors entering (23) in terms of $\mathbf{Y}_{lm}^{(E)}$ and the longitudinal vector:

$$\begin{aligned} \mathbf{Y}_{l,l-1,m}(\hat{\mathbf{k}}) &= \sqrt{\frac{l+1}{2l+1}} \mathbf{Y}_{lm}^{(E)}(\hat{\mathbf{k}}) \\ &+ \sqrt{\frac{l}{2l+1}} \hat{\mathbf{k}} Y_{lm}(\hat{\mathbf{k}}), \\ \mathbf{Y}_{l,l+1,m}(\hat{\mathbf{k}}) &= \sqrt{\frac{l}{2l+1}} \mathbf{Y}_{lm}^{(E)}(\hat{\mathbf{k}}) \\ &- \sqrt{\frac{l+1}{2l+1}} \hat{\mathbf{k}} Y_{lm}(\hat{\mathbf{k}}). \end{aligned} \qquad (24)$$

Substituting (24) in (23) (the transversality condition for the radiation field has to be taken into account) leads to the equivalent expression for $V_l^{(E)}$,

$$V_l^{(E)} = \left( \left( \sqrt{\frac{l+1}{2l+1}} A_l^{(l-1)} + \sqrt{\frac{l}{2l+1}} A_l^{(l+1)} \right) \cdot \left( \mathbf{e}^* \cdot \mathbf{Y}_l^{(E)}(\hat{\mathbf{k}}) \right) \right), \qquad (25)$$

which clearly shows that this term determines the contribution of the electric $2^l$-pole photon to the radiation amplitude $V_{fi}$.

Thus, the series (22) found by us, where $V_l^{(M)}$ is determined by the equation (21) and $V_l^{(E)}$ – by the equation (25), really represents the total amplitude of photon radiation in the direction $\hat{\mathbf{k}}$ with the polarization $\mathbf{e}$ as the sum of the radiation amplitudes of $El$ and $Ml$-photons. Such interpretation gives the additional argumentation to the name "multipole expansion of the radiation amplitude": the expansion in the contributions of the photons of different multipolarity. One can also prove that the irreducible tensor nature of the coefficients $A_{LM}^{(l)}$ (12), (14) of the multipole expansion leads to the angular momentum selection rule: the tensor $A_{LM}^{(l)}$ is only nonzero provided $|J_f - J_i| \le L \le J_i + J_f$, where $J_i$ and $J_f$ are the angular momentum quantum numbers in the initial and final states of the radiating system. Moreover, the parity selection rule also arises: $P_f = (-1)^{l+1} P_i$. The given selection rules just correspond to the angular momentum and parity conservation laws in the radiation of $El$ or $Ml$-photons. Let us also note that in the long-wave approximation $ka \ll 1$ (here $a$ determines the linear dimensions of the radiating system) expanding the spherical Bessel function (4) in the Taylor series and confining ourselves to the first term,

$$g_l(x) = \frac{x^l}{(2l+1)!!},$$

we find from (12) or (14) that $A_{LM}^{(l)}$ has the order $(ka)^l$. Furthermore, in the first nonvanishing order of the long-wave approximation $A_{lM}^{(l)}$ proves to be proportional to well known in the theory of multipole radiation [6,7] magnetic $2^l$-pole moment of transition and $A_{lM}^{(l-1)}$ – to electric one.

Note finally that having eliminated the spherical vector $\mathbf{Y}_{l,l+1,m}(\hat{\mathbf{k}})$ from (23) one can represent the radiation amplitude of $El$-photon $V_l^{(E)}$ (25) entering the multipole expansion (22) in more simple form. It is sufficient for that to express the vector $\mathbf{Y}_{lm}^{(E)}(\hat{\mathbf{k}})$ from the first identity (24), to substitute it in (25) and to use again the transversality condition of the radiation field. As a result we derive the equivalent expression for $V_l^{(E)}$:

$$\begin{aligned} V_l^{(E)} &= \left( D_l \cdot \left( \mathbf{e}^* \cdot \mathbf{Y}_{l,l-1}(\hat{\mathbf{k}}) \right) \right) \\ &= \left( D_l \cdot \left\{ \mathbf{e}^* \otimes Y_{l-1}(\hat{\mathbf{k}}) \right\}_l \right), \end{aligned} \qquad (26)$$

where

$$D_{lM} = A_{lM}^{(l-1)} + \sqrt{\frac{l}{l+1}} A_{lM}^{(l+1)}. \qquad (27)$$

As appears from the above, the tensor $D_{lM}$ (27) has the order $(ka)^{l-1}$ and is proportional to the electric $2^l$-

pole moment of transition in the first nonvanishing order of the long-wave approximation. The next term in the expansion of $D_{lM}$ has the order $(ka)^{l+1}$ and, as it can be proved, is proportional to so-called toroid $2^l$-pole moment of transition, which is determined by the expression for $2^l$-pole toroid moment of classical current $\mathbf{j} = \mathbf{j}_{fi}$ introduced by Dubovik and Cheshkov [4].

## IV. CONCLUSIONS

In the present paper making use of mathematical technique of irreducible tensors the effective, in our opinion, multipole expansion for the probability amplitude of spontaneous radiation (1) has been derived. The expansion is written down in the form of the series (15), (18) or (22). The last form (22) as the sum of the radiation amplitudes of electric $2^l$-pole photons $V_l^{(E)}$ (26) and magnetic $2^l$-pole photons $V_l^{(M)}$ (20) is the most convenient for different applications.

The multipole series contains of course the infinite number of terms, but in case of transitions between the states of the radiating system with definite values of angular momentum and parity the angular momentum and parity selection rules keep only few terms in the series. Note that the interference terms of the form $\text{Re}(V_l^{(M)*}V_{l'}^{(E)})$, $\text{Re}(V_l^{(M)*}V_{l'}^{(M)})$ and $\text{Re}(V_l^{(E)*}V_{l'}^{(E)})$ arising in the expression for the probability of radiation, which contain the products of the amplitudes of different multipolarities allowed by the selection rules, disappear in the total radiation probability. Really, after summation over two independent photon polarizations the products of two different spherical vectors (see (20), (25) and (16)), integrated over all directions of $\hat{\mathbf{k}}$, turn into zero in consequence of orthogonality condition for the spherical vectors. The interference terms will only contribute to the angular distribution (and polarization) of the radiation provided that the initial state of the radiating system is not spherically symmetric, for example, the radiating atom is spin-polarized. For spherically symmetric initial state the angular distribution of the radiation is also spherically symmetric and thus only differs by the multiplier $(4\pi)^{-1}$ from the total radiation probability in which the interference of the amplitude is not manifested.

The found here multipole expansion was used by us in the theoretical study of peculiarities of multipole radiation of light by spin-polarized (i.e., oriented and aligned) atoms in long-wave $ka \ll 1$ approximation [5]. To analyze the long-wave approximation it is convenient to write down the series (22) in the form

$$V_{fi} = V_1^{(E)} + \sum_{l=1}^{\infty}\left(V_l^{(M)} + V_{l+1}^{(E)}\right),$$

where the terms in the parenthesis of the sum over $l$ have the order $(ka)^l$ and $V_1^{(E)}$ has the order $(ka)^0$. The term $V_1^{(E)}$ determines the amplitude of dipole radiation, $V_1^{(M)}$ is the amplitude of magnetic dipole radiation, $V_2^{(E)}$ is the amplitude of quadrupole radiation and so on. In the first nonvanishing order of the long-wave approximation the series will contain no more than two terms owing to the selection rules, and therefore $V_l^{(M)}$ and $V_{l+1}^{(E)}$ amplitudes which have, generally speaking, the same order can only interfere. It has been shown by us [5] that in accordance with above-mentioned the interference of the amplitudes is manifested only provided the so-called state multipoles [10] specifying the orientation and alignment of the radiation atom are nonzero, i.e., the atom is spin-polarized.

Note in conclusion that the derived here multipole expansion can prove to be useful in the study of photoeffect outside the dipole approximation and also in the theory of multiphoton transitions. In the last case the expressions of the form (1) enter as the structural elements the so-called composite matrix elements [11]. We would like to hope that the multipole expansion will find an application for solving outlined here and other problems that arise in studying electromagnetic field-quantum system interaction.

## APPENDIX A: CURRENT DENSITY OF TRANSITION

In case of relativistic electron the expression for the current density vector of transition in the radiation amplitude (1) has the following form (see [6,7]):

$$\mathbf{j}_{fi} = ec\Phi_f^* \hat{\vec{\alpha}} \Phi_i, \qquad (A1)$$

where $e$ is the electron charge, $\Phi_{i,f}$ is the wave function (the Dirac bispinor) of the initial and final states,

$$\hat{\vec{\alpha}} = \begin{pmatrix} 0 & \hat{\vec{\sigma}} \\ \hat{\vec{\sigma}} & 0 \end{pmatrix},$$

$\hat{\vec{\sigma}}$ – are the Pauli spin matrices. For the nonrelativistic electron the expression (A1) turns into following one [7]:

$$\mathbf{j}_{fi} = -\frac{ie\hbar}{2m}\left(\Psi_f^* \nabla \Psi_i - \Psi_i \nabla \Psi_f^*\right) + c\nabla \times \Psi_f^* \hat{\vec{\mu}} \Psi_i. \qquad (A2)$$

Here $\hat{\vec{\mu}} = \frac{e\hbar}{2mc}\hat{\vec{\sigma}}$ is the intrinsic magnetic moment operator of the electron, $m$ is the electron mass, $\Psi_{i,f}$ is the wave function (the spinor taking into consideration the electron spin) of the initial and final states.

One can also derive the equation (A2) for the current density vector of transition in nonrelativistic approximation making use of the expression for the operator of nonrelativistic electron-electromagnetic field interaction in the form

$$-\frac{e}{2mc}\left(\hat{\mathbf{p}}\cdot\hat{\mathbf{A}} + \hat{\mathbf{A}}\cdot\hat{\mathbf{p}}\right) + \frac{e^2}{2mc^2}\hat{\mathbf{A}}^2 - \left(\nabla\times\hat{\mathbf{A}}\right)\cdot\hat{\vec{\mu}}, \quad (A3)$$

where $\hat{\mathbf{p}} = -i\hbar\nabla$ is the momentum operator of electron and $\hat{\mathbf{A}}$ is the vector potential of quantized electromagnetic field [6]. In the first order of perturbation theory the quadratic in $\hat{\mathbf{A}}$ term entering (A3) does not contribute, and the known expression for spontaneous radiation probability [6] (see Sec. II of the paper) is derived, where the current density vector of transition entering the probability amplitude (1) is determined by the formula (A2). The expression (A2) for the current density vector of transition is obviously generalized in case of $N$-electron system:

$$\mathbf{j}_{fi}(\mathbf{r}) = \sum_{a=1}^{N}\int\left[-\frac{ie\hbar}{2m}\left(\Psi_f^*\nabla_a\Psi_i - \Psi_i\nabla_a\Psi_f^*\right) + c\nabla_a\times\Psi_f^*\hat{\vec{\mu}}_a\Psi_i\right]\delta(\mathbf{r}-\mathbf{r}_a)d\mathbf{r}_1 d\mathbf{r}_2..d\mathbf{r}_N.$$

## APPENDIX B: CHANGING THE COUPLING SCHEME IN THE SCALAR PRODUCT

One can change the coupling scheme in the scalar product $\left(\{A_L \otimes B_l\}_1 \cdot \mathbf{a}\right)$ without 6$j$-symbols. Let us write down the scalar product, taking into account the definitions (6) and (7), in the form of the sum:

$$\left(\{A_L \otimes B_l\}_1 \cdot \mathbf{a}\right) = \sum_{m,m_1,m_2} C_{Lm_1 lm_2}^{1m} A_{Lm_1} B_{lm_2} a_{-m}(-1)^{-m}. \quad (B1)$$

Making further use of the symmetry property for the Clebsh-Gordan coefficient,

$$C_{Lm_1 lm_2}^{1m} = (-1)^{l+m_2}\sqrt{\frac{3}{2L+1}}C_{1,-m lm_2}^{L,-m_1},$$

and substituting this identity in (B1), where $m = m_1 + m_2$, we find the sought relation:

$$\left(\{A_L \otimes B_l\}_1 \cdot \mathbf{a}\right) = (-1)^l\sqrt{\frac{3}{2L+1}} \times \left(A_L \cdot \{\mathbf{a}\otimes B_l\}_L\right). \quad (B2)$$

Note that the relation (B2) with $L = l = 1$ turns into obvious identity for the scalar triple product of the vectors.

___________________________________________